\documentclass[11pt, subeqn]{article} 
\usepackage{amsmath}
\usepackage{amsfonts}
\usepackage{amssymb}
\usepackage{graphics}
\usepackage{graphicx}
\usepackage{amssymb}
\usepackage{color,hyperref}
\usepackage{setspace}
\definecolor{darkgreen}{rgb}{0.0,0.4,0.0}
\hypersetup{colorlinks,breaklinks,linkcolor=darkgreen,urlcolor=darkgreen,anchorcolor=darkgreen,citecolor=darkgreen}
\newcommand{\be}{\begin{equation}}
\newcommand{\ee}{\end{equation}}
\newcommand{\ba}{\begin{align}}

\newcommand{\ra}{\rightarrow}

\newcommand{\mn}{\mu\nu}

\newcommand{\al}{\alpha}
\newcommand{\op}{\oplus}
\newcommand{\lb}{\lambda}
\newcommand{\mK}{\mathcal{K}}
\newcommand{\om}{\ominus}
\newcommand{\kp}{\kappa}
\newcommand{\mN}{\mathcal{N}}
\newcommand{\mW}{\mathcal{W}}
\usepackage[dvips,letterpaper,margin=0.8in]{geometry}
\title{
\huge{Gamma Ray Bursts, \\The Principle of Relative Locality\\ {\Large and} \\Connection Normal Coordinates}\\[2cm]
\Large{          Anna McCoy}\\\small{\emph{Perimeter Scholars International}}\\[.7cm]
{\tiny supervised by}\\[.7cm]
{\Large Lee Smolin}\\
\emph{Perimeter Institute of Theoretical Physics}\\
\emph{31 Caroline Street North}\\
\emph {Waterloo N2L 2Y5, Ontario, Canada}\\[4cm]
A thesis submitted in partial fulfillment of the requirements for the Degree of \\
MASTERS OF SCIENCE\\ [1cm]
University of Waterloo, June 2, 2011
}

\author{}\date{}
\begin{document}
\maketitle
\thispagestyle{empty}
\newpage
\pagenumbering{roman}
\newpage
\begin{center}Abstract\end{center}
The launch of the Fermi telescope in 2008 opened up the possibility of measuring the energy dependence of the speed of light by considering the time delay in the arrival of  gamma ray bursts emitted simultaneously from very distant sources.The expected time delay between the arrival of gamma rays of significantly different energies as predicted by the framework of relative locality has already been calculated in Riemann normal coordinates.  In the following, we calculate the time delay in more generality and then specialize to the connection normal coordinate system as a check that the results are coordinate independent.   We also show that this result does not depend on the presence of torsion.

\tableofcontents
\newpage 

\pagenumbering{arabic}
The launch of the Fermi telescope in 2008 opened up the possibility of measuring the energy dependence of the speed of light by considering the time delay in the arrival of  gamma ray bursts.
To first order, the energy-dispersion relationship arising from an energy dependent speed of light common to many theories of quantum gravity, is expected to have the form
\be
E^2-p^2+\alpha \frac{E^3}{m_{QG}}=m^2,
\ee
where $m_{QG}$ is the mass scale of quantum gravity and $\alpha$ is a proportionality constant.  From this we would expect a time delay proportional to $\frac{\Delta E }{m_{QG}}L$ where L is the distance traveled \cite{exp}.  For shorter distances, the time delay is approximately zero.  However, for cosmological distances we expect there to be a measureable time delay.  Although the quantum gravity effects are entangled with possible cosmological phenomena, using statistical analysis of the probability of two gamma rays being emitted at the same time, it may be possible to separate quantum effects from cosmological effects and actually measure the quantum gravity constant. Although we do not assume that $m_{QG}$ is the Planck mass, we expect that it is of similar magnitude.  The expected time delay between the arrival of gamma rays of significantly different energies as predicted by the framework of relative locality \cite{rel} is calculated in the Riemann normal coordinate system in \cite{Gamma}.  In the following, we generalize the time delay calculation, discuss both first and second order effects and then specialize to the connection normal coordinate system.   The results in this paper are consistent with the time delay calculated in \cite{Gamma}, thus confirming that the time delay is independent of the coordinate system.  We also show that this result does not depend on the presence of torsion to first order.    
\section{The Planck Scale Puzzle}
The role of Planck scale in physics is not clearly understood.  The fundamental constant of quantum mechanics, $\hbar$, first appeared in Planck's derivation of blackbody radiation \cite{planck} and was later linked explicitly to energy quanta emitted and absorbed in Einstein's paper on the photoelectric effect \cite{einstein1}.  It was during this same period that Lorentz, in an attempt to explain the null results of the the Michelson-Morley experiment and the constant speed of light in Maxwell's theory, developed the Lorentz transformations that became the basis of Einstein's special relativity \cite{einstein2}.  While the theory of special relativity allowed for a constant speed of light, the upper bound on velocity contradicted the instantaneous effects of gravity in Newton's gravity.  To unify the two theories, Einstien once again adapted the concept of space-time by introducing curvature giving rise to general relativity. There still remains, however, a theory that is not compatible with general relativity: quantum mechanics.   \\\\
It is no surprise that as we seek to reconcile quantum mechanics and general relativity, we expect some combination of the constants, $c$, $\hbar$ and $G_N$, which form the Planck scale, to play a role. Much like the constant speed of light did prior to special relativity, an invariant smallest length scale contradicts the Lorentz invariance.   While there is currently little evidence to support the existence of a smallest length, it appears in many of the modern theories; a smallest length scale appears in string theories as the minimal length of the string and in quantum field theories as ultraviolet cutoffs.  Another important indication that there is a fundamental length scale comes form our inability to probe scales smaller than the Planck scale.  The amount of energy necessary to probe such a length scale would have a Schwarzschild radius of approximately the Planck length. Thus, if we try to probe distances on the order of the Planck length or smaller, the high amount of energy necessarily concentrated in that small of a distance would create a black hole and no information could be recovered.  At this point, the current concept of space-time no longer has any meaning. \\\\
Over the years, a number of theories have been proposed to answer the Planck scale puzzle.  The particular class of theories in which we are interested is commonly referred to as Doubly Special Relativity (DSR) characterized by the following properties:
\begin{itemize}
\item Observer independence is preserved
\item Two constants are invariant under modified Lorentz transformations: the speed of light and some constant corresponding a fundamental length or mass scale.
\end{itemize}
The first DSR theory was proposed in 2000 by Amelino-Camelia\cite{DSR1}, 
followed shortly by a different theory by Smolin and Magueijo using a different basis \cite{2DSR}.  It quickly became evident that there was an entire class of DSR theories where the different formulations of the theory correspond to the choice of coordinate system on momentum space--the natural %
setting for DSR \cite{DSR2}\cite{DSR3}. 
\\\\
Some DSR theories  incorporate a fundamental length or mass scale by deforming the Poincar\'e transformations.  While the rotations and spacial operators remain unchanged, the boost must be deformed to accommodate an invariant length.  The resulting deformed transformations form a 10-dimensional algebra referred to as the $\kappa$-Poincar\'e algebra \cite{DSR3}.   It can be shown that the action of the deformed boosts on momentum space necessitates that the space-time arising from the phase space algebra is non-commutative \cite{Kpoincare} and takes the form suggested by Snyder  as a possible solution to the Planck scale problem \cite{snyder}.  It was also Snyder who first observed that curved momentum space would necessarily lead to non-commutative space-time.  The idea that curved momentum space might be the correct arena in which to formulate a consistent theory of quantum gravity is not a new one.  It was first proposed by Born in his famous paper on the reciprocity of space-time and momentum \cite{reciprocity} in 1938.   This idea of reciprocity was furthered by Majid, where he showed that under non-abelian Fourier transforms any noncommutative construction on space-time can be mapped to the non-Abelian (curved) momentum space \cite{QGroup1}.  This framework is the basis for a number of the quantum group approaches to DSR.\\
\\
Although the DSR theory has managed to avoid many of the constraints faced by theories that accommodate an addition fundamental scale by breaking Lorentz invariance through the existence of a preferred reference frame, there are several notable problems that must still be addressed.  One of the major issues is the difficulty in recovering the macroscopic behavior, often referred to as the Soccer Ball Problem.   The non-linear addition of momentum is expected to have correction to first order on the order of $\frac{E}{m_p}$ where $E$ is the energy of the object.  We expect this relationship to hold for macroscopic objects such as soccer balls as well as microscopic particles.  However,  $E\sim m_{ball}$, $m_p\sim10^{-5}g$ and $m_{ball}\sim 500g$.  Thus the first order correction term would be $\sim 10^7$g, an thus clearly observable at the macroscopic level.  Another issue, raised by Sabine Hossenfelder, is that the simultaneous conditions of no preferred frame of reference and two invariant scales leads to non-local interactions \cite{problem}.  It is this problem that  motivated the development of the theory of Relative Locality \cite{rel}.  In addition to addressing the issue of non-local effects, Relative Locality also provides a solution to the Soccer Ball Problem \cite{soccer}.

\section{The Principle of Relative Locality}
The approach taken by Amelino-Camelia, Freidel, Kowalski-Glikman and Smolin in [2]  is that the paradox of non-local interactions is simply a consequence of a deeper and more fundamental structure underlying space-time \cite{rel}.  They propose that locality is not a universal idea, but rather a relative concept.   The  principle of relative locality states :
\begin{quote}
\emph{Physics takes place in phase space and there is no invariant
global projection that gives a description of processes in space-time. From their measurements local observers can construct descriptions of particles moving and interacting in a spacetime, but different observers construct different spacetimes, which are observer-dependent slices of phasespace.}
\end{quote}
At first this may seem like a preposterous idea.  After all, we observe that interactions are local.  However, what reason do we have to suppose that on the cosmological scale the space-time that we impose around us is the same as that light-years away? Why, after all of the modifications to space and time in the development of modern physics do we still cling so tightly to the notional that space-time is in some way universal?  As Amelino-Camelia et al. points out, the concept of space-time is constructed using the exchange of light signals measuring only the time of flight and neglecting the energy of the light signals.  But should we expect the energy of the light signals to remain unimportant in the high energy limit?  Although the familiar geometry of space time must be recovered in the low energy limit, there is no reason to assume that the energy is negligible in the construction of the quantum geometry.   It turns out that  absolute locality is equivalent to the assumption that momentum space is linear \cite{rel}.  In the framework of relative locality, we do not ignor the energy of the light signals in constructing space-time.   Instead each observer sees a different space-time that depends on energy and momentum.  If an observer is near the interaction, then the observer sees the interaction as local.  However, if the observer is distant from the interaction, then the interaction may appear non-local.  This is not a physical non-locality, but rather due to the geometry of momentum space.  More specifically, as we will see, this apparent non-locality is due to the non-metricity of momentum space.  \\\\
The framework in which we are working is the semiclassical regime where $\hbar$ and $G_N$ can both be neglected but the ratio
\begin{equation}
\sqrt{\frac{\hbar}{G_N}}=m_p
\end{equation}
is held fixed.  Note that we are working in units where $c=1$.  Because we are taking $G_N\ra0$ and $\hbar\ra0$, the Planck length $l_p=\sqrt{\hbar G_N}\ra0$.   Although we expect effects from the general relativity and quantum space-time geometry to be negligible, there are new phenomena occurring at the scale defined by $m_p$.  
\subsection{Space-time and the Dynamics of Relative Locality}
The action of a particle in the relative locality theory given in \cite{rel} is
\be
S^{total}=\sum_jS^j_{free}+S^{int}.
\ee
The free action for incoming particles is given by 
\be
S^j_{free}=\int_{-\infty}^0 ds(x^a_j\dot{k}^j_a+\mathcal{L}_j\mathcal{C}^j(k)),
\ee
and the free action for outgoing particles is 
\be
S^j_{free}=\int_0^\infty ds(x^a_j\dot{k}^j_a+\mathcal{L}_j\mathcal{C}^j(k)), 
\ee
 where $\mathcal{L}$ is the Lagrange multiplier imposing the mass shell condition $C^j(k)\equiv D^2(k)-m^2_j$.  We define $D^2(k)\equiv D^2(k,0)$ to be the geometric distance from the origin to a point $k$ in the momentum manifold $\mathcal{P}$.    This distance can be physically interpreted as $D^2(k,0)=m^2,$ where $m$ is the mass of the particle.  
The interaction term is given by 
\be
S^{int}=\mathcal{K}(k(0))_az^a,
\ee
where $\mathcal{K}$ is the composition rule for the particle intereactions.\\\\
There are two types of space-time coordinates in the action, $x^a$ and $z^a$.  The first kind, $x_a$ arises as the conjugate of the momentum coordinates $k_a$ satisfying $\{x^a,k_b\}=\delta^a_b$.   From this we see that $x_a$ coordinatizes the cotangent space $T^*(p)$. Because the momentum space is curved, the worldlines of particles with different momentum live in different cotangent spaces.  So, in order to be able to talk about particles interacting, we will need to parallel transport the information about the particles to the cotangent plane at the origin, which is coordinatized by the other type of space-time coordinates, $z_a$.  We will refer to the $z_a$ coordinates as the interaction coordinates.  Unlike the conjugate coordinates, the interaction coordinates do not correspond to a physical momentum.  In the Lagrangian, they appear as a Lagrange multiplier enforcing the conservation law at interactions.  We will see later that they play a very important role in dealing with particle interactions.  \\\\
The relationship between the conjugate spacetime coordinates and the interaction coordinates is given by the equations of motion. By varying the total action and integrating by parts, we find the following relationships for the free part of the action:
\be
\dot{k}^j_a=0\qquad \dot{x}_j^a=\mathcal{L}_j\frac{\delta C^j}{\delta k_a}\qquad C^j(k)=0.
\ee
From the boundary terms and interaction terms we find that 
\be
\label{relation}\mathcal{K}(k(0))_a=0,
\ee
which gives us the energy-momentum conservation law at each interaction.  We also have that  
\be
x_j^a(0)=z^b\frac{\delta\mathcal{K}_b}{\delta k^j_a},
\ee
which relates the space-time coordinate at the ends of the worldline to the interaction coordinates by the specific choice of momentum conservation law imposed on the interaction of the particles.    

\section{Geometry of Momentum Space}
Once we relax the assumption that the momentum space is linear, we will need to define a composition rule for the four-momentum.  In the following section we briefly summerize the geometry of the non-linear momentum space. A more complete description can be found in \cite{Geometry}.  To combine momenta we  define a $C^\infty$ map 
\begin{align}
\begin{array}{cc}
\op:\mathcal{P}\times\mathcal{P}\ra\mathcal{P}\\
(p,q)\ra(p\op q)
\end{array}
\end{align}
to be a left invertible composition law that has the following properties:\footnote{from these properties we see that momentum composition forms an algebra.}  
  \begin{enumerate}
  \item There is a unit element $0$ such that $(0\op p)=p=(p\op 0)$
  \item The map has an inversion $\ominus :P\ra P$ such that 
  \begin{equation*}
  (\om p\op p)=0
  \end{equation*}
   and 
   \begin{equation*}p\op(\om p\op q)=q=\om p\op(p\op q).
   \end{equation*}
 \end{enumerate}

Using the composition rule we can define a left multiplication operator
\begin{equation}
\begin{array}{cc}
\L_p:\mathcal{P}\ra\mathcal{P}\\
L_p(q)\equiv (p\op q)
\end{array}
\end{equation}
that statisfies the identity and inverse conditions 
\begin{equation}
 L_p(0)=p\qquad\mathrm{and}\qquad L_p^{-1}=L_{\om p}.
\end{equation}
Since the composition of momenta is non-linear, there is no reason to suppose that the composition should be associative or commutative.  In fact, it is the lack of associativity that gives rise to the curvature of momentum space and the lack of commutativity that measures the torsion.  Thus, it is convient to also define a right multiplication operator
\begin{equation}
R_p(q)=(q\op p).
\end{equation}
In this paper we only consider momenta in the neighborhood of the origin.  However, if we want to consider momenta away from the origin, we can define the translated composition law
\begin{equation}
p\op_kq\equiv L_k\big(L_k^{-1}(p)\op L_k^{-1}(q)\big),
\end{equation}
 where all of the previous properties hold with $0_k=k$ as the new identity.  Using these rules for composition we can now define a conservation law $\mK$ that enforces conservation of energy and momentum.  For example, the conservation law of a two particle interaction would  look like
 \be
 \mK=p\op q=0,
 \ee
 which is satisfied by $q=\om p$.  For three particle interactions, the composition is not quite so trivial.  If we have incoming momenta $p,q$ and outgoing momentum $r$ then the composition law could be 
 \be
 \mK=(p\op q)\om r,  \qquad\mathrm{or}\qquad \mK=p\op(q\om r),\qquad \mathrm{or}\qquad \mK=(q\op p)\om r
 \ee
 or any one of the 12 possible composition laws that, in the presence of torsion and curvature, are all unique.  
 
 For small momenta, there is a useful way to expand the composition rule.  If we take $q$ to be small, then we can write
 \be
 p\op q=p_a+(U_p^0)_a^bq_b\label{aaa}.
 \ee
 That is, to compose $p$ and $q$, we have to first parallel transport the momentum vector $q$ from the origin to $p$ and then sum them.   So, in order to make sense of this composition law, we will need to define a parallel transport operator.

\subsection{Parallel Transport}
We define the left parallel transport operator from the tangent space at q, denoted $T_q\mathcal{P}$ to the tangent space at $p\op q$, denoted $T_{p\op q}\mathcal{P}$, as 
\begin{equation}
(U^q_{p\op q})^b_a\equiv\frac{\partial(p\op q)_a}{\partial p_b}=(d_qL_p)^b_a,
\end{equation}
where $d_qL_p$ is the differential of $L_p$ at the point $q$.  The right parallel transport is similarly defined as 
\begin{equation}
(V^p_{p\op q})^b_a\equiv\frac{\partial(p\op q)_a}{\partial p_b}=(d_pR_q)^b_a.
\end{equation}
The parallel transport operator for the inverse is  
\begin{equation}
(I^p)^b_a=\frac{\partial(\om p)_a}{\partial p_b}.\label{bb}
\end{equation}
The inverses are given by
\be
\left((U^q_{p\op q})^b_a\right)^{-1}=(U_q^{p\op q})^b_a\qquad \left((V^p_{p\op q})^b_a\right)^{-1}=(V_p^{p\op q})^b_a\qquad \left((I^p)^b_a\right)^{-1}=(I^{\om p})^b_a
\ee
In general, the parallel transport operator from the origin to $p$ can be written as  
\ba
(U^q_p)_a^c=\mathcal{P}\exp\left(-\int_{a_0}^a\Gamma^{bc}_a(t)\dot{\gamma}_b(t)dt\right)q_c\label{1a}
\end{align}
where $\gamma(t)$ is a parameterized curved from the origin to p,  $\gamma(a_0)=q$, $\gamma(a)=p$,and $\Gamma^{bc}_a(t)$ is the connection coefficient parameterized by $t$.  We can  expand the parallel transport operator as 
\begin{equation}
U^0_p=\delta^b_a-\sum_{n\geq1}\frac{1}{n!}\Gamma^{\al_1...\al_nb}_ap_{\al_1}...p_{\al_n},\label{expand}
\end{equation}
where 
\begin{equation*}
\Gamma^{\al_1...\al_nb}_a=\partial^{\al_a}\Gamma^{\al_2...\al_nb}_a-\Gamma^{\al_1\al_i}_\sigma\Gamma^{\al_1...\al_{i-1}\sigma\al_{i+1}...\al_nb}_a.
\end{equation*}
The connection coefficients are given by\footnote{The usual convection is that the connection is given by
\be
\nabla_\nu T^{\lb_1...\lb_p}_{\mu_1...\mu_q}=\partial_\nu T^{\lb_1...\lb_p}_{\mu_1...\mu_q}+\Gamma^{\lb_1}_{\nu\kp} T^{\kp...\lb_p}_{\mu_1...\mu_q}+...+\Gamma^{\lb_p}_{\nu\kp}T^{\lb_1...\kp}_{\mu_1...\mu_q}-\Gamma^{\kp}_{\nu\mu_1}T^{\lb_1...\lb_p}_{\kp...\mu_q}-...-\Gamma^\kp_{\nu\mu_q}T^{\lb_1...\lb_p}_{\mu_1...\kp}
\ee
but keeping with the idea that the momentum coordinates have upper indices we will rewrite this as 
\be
\nabla^\nu T^{\lb_1...\lb_p}_{\mu_1...\mu_q}=\partial^\nu T^{\lb_1...\lb_p}_{\mu_1...\mu_q}+\Gamma^{\nu \kp}_{\mu_1}T^{\lb_1...\lb_p}_{\kp...\mu_q}+...\Gamma^{\nu\kp}_{\mu_q}T^{\lb_1...\lb_p}_{\mu_1...\kp}-\Gamma^{\nu \lb_1}_\kp T^{\kp...\lb_p}_{\mu_1...\mu_q}-...-\Gamma^{\nu\lb_p}_\kp T^{\lb_1...\kp}_{\mu_1...\mu_q}. 
\ee

}
\begin{equation}
\left.\Gamma^{cb}_a(p)\equiv-\frac{\partial}{\partial r_c}\frac{\partial}{\partial q_b}(r_a\op q_a)\right|_{r_a,q_a=p_a.}
\end{equation}
The antisymmetric part of the connection coefficient is defined as the torsion.  That is
\be
T^{cb}_a(p)=\frac{1}{2}\Gamma^{[cb]}_a(p)=\left.-\frac{\partial}{\partial r_c}\frac{\partial}{\partial q_b}\big((r_a\op q_a)-(q_a\op r_a)\big)\right|_{r_a,q_a=p_a.}
\ee
To first order, 
\be
(U_p^0)_a^b=\delta_a^b-\Gamma^{cb}_ap_c\qquad \qquad (V_p^0)_a^b=\delta_a^b-\Gamma^{bc}_ap_c \label{bbb},
\ee
where the expansion of of $V$ is determined its relation it to $U$.  
Using \eqref{aaa} and \eqref{bbb},  we can expand the composition of small momenta to leading order as
\begin{equation}
(p\op q)_a=p_a+q_a-\Gamma^{bc}_a(0)p_bq_c.  
\end{equation}
By applying the condition that $\om p\op p=0$, we have that 
\begin{equation}
(\om p)_a=-p_a-\Gamma^{bc}_a(0)p_bp_c.
\end{equation}
The expansion to second order is slightly less trivial, but from \eqref{1a} we have that 
\ba
U^q_p&=\delta^b_a-\Gamma^{cb}_a(p_c-q_c)-\frac{1}{2}\partial^b\Gamma^{dc}_a(p_bp_d-q_bq_d)+\frac{1}{2}\Gamma^{be}_a\Gamma_e^{dc}(p_dp_b+q_bq_d)-\Gamma^{be}_a\Gamma_e^{dc}p_bq_d\\
V^q_p&=\delta^b_a-\Gamma^{cb}_a(p_b-q_b)-\frac{1}{2}\partial^c\Gamma^{bd}_a(p_bp_d-q_bq_d)+\frac{1}{2}\Gamma^{ec}_a\Gamma^{db}_e(p_bp_d+q_bq_d)-\Gamma^{ec}_a\Gamma^{db}_ep_bq_d\\
\end{align}
where $\Gamma$ and $\partial \Gamma$are evaluated at zero and the partial derivative $\partial^b\Gamma^{dc}_a$ is taken with respect to which ever momentum the derivative is contracted with.  That is $\partial^b\Gamma^{dc}_ar_b=\partial_r^b\Gamma^{dc}_ar_b$.  
\subsection{The parallel transport operator for $\om$}
There is a third type of parallel transport operator given by \eqref{bb} that is the operator that takes $p$ to $\om p$.  This operator has some curious properties that deserve some discussion.  
In \cite{Gamma} the identity
 \be
 U^p_0=-V^{\om p}_0I^p \label{gm1}
 \ee
 is derived as a special case of the identity 
 \be
 V_q^{\om p}I^p=-U^{p\op q}_qV^p_{p\op q}\label{z},
 \ee
 which follows from the observation that $0=\partial_p(\om p\op(p\op q)).$  
 However, note that if we apply the inverse of $ V_q^{\om p}$ we have that 
 \be
 I^p= -V^q_{\om p}U^{p\op q}_qV^p_{p\op q}.
 \ee
 That is, there is an arbitrary $q$ in the equation for $I^p$.  This seems to imply a path independence of $I^p$.  This leads to the question of what does it mean to parallel transport between two points not at the origin?  However, we do have some notion of what it means to parallel transport from a point to the origin,  If we choose the path between two points to be the geodesic from the first point to the origin and then the origin to the second point, this is equivalent to parallel transporting both points to the origin to the interaction plane where their relation can easily be determined.  In this case we can write 
\be
U_p^q=U_p^0U_0^q.\label{idpath}
\ee
This choice of path is then consistent with the taylor expansion of the parallel transport operator in \eqref{expand}, where by expanding the parallel transport operator about the origin, we implicitly chose this path.  

\subsection{Curvature\label{curved}}
With the second order expansion, we to see curvature play a role.  Indeed, when we consider a closed curve created by, for example, $U_0^pV_p^0$ or, alternatively, $V_0^pU_p^0$.  Note that these two paths do not generate the same curvature.  To see this we can calculate to second order
\be
(V_0^pU_p^0)_a^f=\delta_a^f+2T_a^{fb}+\frac{1}{2}(\partial^f\Gamma^{bd}-\partial^b\Gamma^{df}+2\Gamma^{eb}_a\Gamma^{[fd]}_e-2\Gamma^{[eb]}_a\Gamma_e^{df})p_bp_d.  
\ee
Thus we will define 
\be (R_{VU})_a^{bdc}=\frac{1}{2}(\partial^c\Gamma^{bd}-\partial^b\Gamma^{dc}+2\Gamma^{eb}_a\Gamma^{[cd]}_e-2\Gamma^{[eb]}_a\Gamma_e^{dc}).
\ee
Similarly,
\be
(U_0^pV_p^0)_a^f=\delta_a^f+2T^{bf}p_b+\frac{1}{2}(\partial^b\Gamma^{df}_a-\partial^f\Gamma^{bd}_a+2\Gamma_a^{be}\Gamma^{[df]}_e-2\Gamma^{[be]}_a\Gamma_e^{fd})p_bp_d
\ee
and so
\be
(R_{UV})_a^{bdc}=\frac{1}{2}(\partial^b\Gamma^{dc}_a-\partial^c\Gamma^{bd}_a+2\Gamma_a^{be}\Gamma^{[dc]}_e-2\Gamma^{[be]}_a\Gamma_e^{cd}).
\ee
Notice that $R_{UV}$ and $R_{VU}$ are not the same.  They differ by 
\be
R^{bdc}_a=\frac{1}{2}(R_{VU}-R_{UV})=\frac{1}{2}(\partial^c\Gamma^{bd}-\partial^b\Gamma^{dc}+2\Gamma^{(eb)}_a\Gamma^{[cd]}_e-2\Gamma^{[eb]}_a\Gamma_e^{(cd)})\label{curve}.
\ee
The fact that the  "curvature" appears to be a consequence of the difference between left and right parallel transport operators suggests that it is directly tied to the presence of torsion.  However, note that even if torsion were zero, the two derivatives would remain, thus it is not just the torsion that plays a role.

\subsection{Non-Linear Phenomena in Gamma Ray Bursts}

The theory of general relativity is formulated on a Riemannian manifold, which is metric compatible.  That is 
\begin{equation}
\nabla^ag^{bc}=0.
\end{equation}
This choice is natural because it preserve the length of a vector that is parallel transported from tangent or cotangent space to another.  However, we do not assume that $\mathcal{P}$ is metric compatible.  In fact, it is the presence of non-metricity, defined as 
\begin{equation}
N^{abc}\equiv\nabla^ag^{bc}.
\end{equation}
that leads to the expected dispersion relationship that creates the time delay exhibited by the gamma rays.  If we allow for non-metricity and torsion, the connection coefficients can then be written as 
\begin{align}
\Gamma^{(ab)}_c=\{^{ab}_c\}+\frac{1}{2}g_{ci}\left(T^{iab}+T^{iba}-N^{abi}-N^{bai}+N^{iab}\right)
\end{align}
Where $T^{iab}=g^{bd}T^{ia}_d$ and $\{^{ab}_c\}$ is the Christoffel symbol.  
 
\section{Time Delay in Gamma Ray Bursts}
\begin{figure}[h]
 \center
 \includegraphics[width=.5\textwidth]{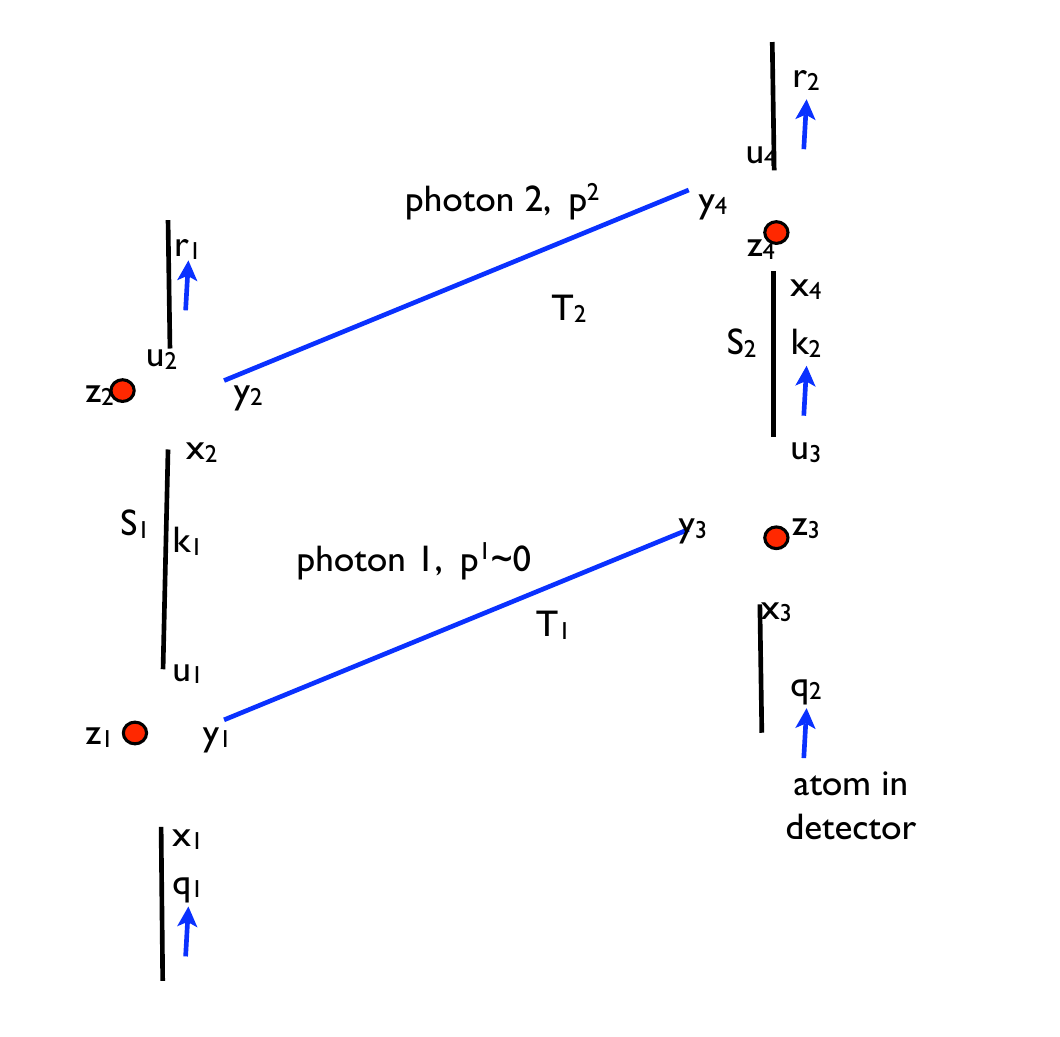}
 \caption{\label{gamma}The set up for the gamma ray burst experiment.} 
 \end{figure}

We now want to calculate the time delay that would be observed in the arrival of two gamma rays of different energies.   Figure \ref{gamma}, taken from \cite{Gamma}, shows the emission of two photons (gamma rays) and their subsequent detection by some detector (Fermi Telescope).  The Fermi Telescope will only detect the  energy of two gamma rays (photons) and the difference in  their arrival times. So, we will need to derive an expression for the time delay in terms of the photon energies.  The final result may depend on the travel time of the photons, but it should be independent of all the variables associated with the detector or the source of the gamma rays.  
 \\\\
In Figure \ref{gamma}, a source with initial momentum $q^1$ emits a photon with momentum $p^1$ and is left with momentum $k^1$.  The emission occurs at the end of the world-line, denoted by $x_1$, of the source with momentum $q^1$.  The beginning of the world-line of the photon with momentum $p^1$ is $y_1$.  The worldline of the source, which now has momentum $k^1$ continues at $u_1$.  The distinction between $u_1$, $y_1$, and $x_1$ must be made because each "particle" has a different momentum and thus lives in  different cotangent spaces.  In this sense,  we can think of $u_1$ as being the beginning of a new world line for the detector.  However, because we want to talk about the interaction that occurred with the emission of the photon, we need to parallel transport everything to the same cotangent space at the origin.  The location of the interaction in the interaction plane is given by $z_1$.  After a proper time $S_1$ the source will emit another photon.  The two photons with momenta $p^1$ and $p^2$ will travel for proper times $T_1$ and $T_2$ respectively before being absorbed by the detector.  The arrival time of the photons differs by a proper time $S_2$.  To calculate the time delay \be
\Delta S\equiv S_2-S_1
\ee
we need some way of relating $S_1$ and $S_2$.  To do this we use the fact that the two photons are emitted and detected by the same source and detector respectively.   That is, both photons must be at coordinate $z_1$ and $z_4$. Intuitively, we would suppose that we could relate the distance that the first photon travels to the distance that the second photon travels by the relationship  
\be
\label{suppose}S_1\dot{z}(k^1)+T_2\dot{z}(p^2)=T_1\dot{z}(p^1)+S_2\dot{z}(k^2),
\ee
where $\dot{z}(r)$ is the velocity of a particle with momentum $r$ as seen by an observer at the interaction plane (at the origin).  We will see later that this is not exactly true, but it is a good starting to try to relate $S_1$ and $S_2$. 
\subsection{Coordinate Independent Time Delay}
The velocity of a particle with momentum $p$ is given by the equations of motion as
\begin{equation}
\dot{x}^a=\mathcal{L}\frac{\delta C}{\delta p_a}.  
\end{equation}
To be consistent with the linear theory relationship between momentum and space-time we take $\mathcal{L}=\frac{1}{2m}$.  
Thus the velocity of a particle with momentum $p$ will be given by
\be
\dot{x}^a=\frac{1}{2m}\frac{\partial (D^2(p)-m^2)}{\partial p_a}=\frac{1}{2m}\frac{\partial D^2(p)}{\partial p_a}.
\ee
However, we cannot add and subtract these velocities because they live in different cotangent planes, so we need to calculate the $\dot{z}$'s, which can then be combined to calculate the time delay.  
From the equations of motion, we also have a way to relate the velocity of the particle at the end of its world-line to the interaction plane given by \eqref{relation}.  
To simplify notation, we define a new quantity
\begin{equation}
(\mathcal{W}_{x_i})^a_b=\pm\frac{\delta \mK_b}{\delta k^i_a},
\end{equation}
which can be thought of as the parallel transport matrix that takes the interaction coordinate in $T_{\mathcal{P}}^*(0)$ to the end of the worldline living in $T_{\mathcal{P}}^*(k)$ . The index $i$ simply denotes which particle we are considering and the + and - correspond to incoming and outgoing particles respectively.  Using the matrices, the conjugate coordinates and the interaction coordinates can be related by
\begin{equation}
z_i(x)=x_i\mW_{x_i}^{-1},\qquad z_i(u)=u_i\mW_{u_i}^{-1},\qquad z_i(y)=y_i\mW_{y_i}^{-1}
\end{equation}
Before we can actually do the calculation we need to choose a specific energy-momentum conservation law for the interactions.  To be consistent with \cite{Gamma} we will choose these to be 
\begin{align}
\label{emc1}\mK^1&=(q^1\om k^1)\om p^1=0\qquad\qquad
\mK^2=(k^1\om r^1)\om p^2=0\\
\label{emc2}\mK^3&=p^1\op(\om k^2\op q^2)=0\qquad\quad
\mK^4=p^2\op(\om r^2\op k^2)=0
\end{align}
Now, we want to calculate the $\dot{z}$'s.  However, there is some ambiguity on how to determine what $\dot{z}$ is.  Does $\dot{z}(k^1)$ correspond to $\dot{x}_2\mW_{x_2}^{-1}$ or $\dot{u}_1\mW_{u_1}^{-1}$ , or some combination of both?  To get around this problem, we follow the same procedure as outlined  in \cite{Gamma} for the connection normal coordinate case.  To start with we define 
\begin{equation}
\dot{x_2}S_1=x_2-u_1.
\end{equation}
However, $x_2-u_1$ does't really have any meaning so we need to parallel transport everything to the interaction plane at the origin using the relationship between the conjugate coordinates and the interaction coordinates.  We then have that 
\begin{equation}
\label{z21}\dot{x_2}\mW_{x_2}^{-1}S_1=z_2-z_1\mW_{u_1}\mW^{-1}_{x_2}.
\end{equation}
In the above expression, the two end points of the world-line of the detector between the emissions of the photons is projected to the interaction plane.  Notice that instead of simply comparing $z_1$ and $z_2$ as we did $x_2$ and $u_1$, the comparison is induced by projecting $z_1$ back to the cotangent plane at $k$, moving along the "straight" geodesic to $x_2$ and then projecting back to the interaction plane. 
Similarly
\begin{align}
\label{z31} \dot{u_3}\mW^{-1}_{u_3}S_2=z_4\mW_{x_4}\mW_{u_3}^{-1}-z_3\\
\label{z32}\dot{y}_3\mW^{-1}_{y_3}T_1=z_3-z_1\mW_{y_1}\mW_{y_3}^{-1}\\
\label{simp1}\dot{y}_4\mW^{-1}_{y_4}T_2=z_4-z_2\mW_{y_2}\mW_{y_4}^{-1}.
\end{align}
Adding \eqref{z31} and \eqref{z32}, we can eliminate $z_3$.
\begin{equation}
\label{simp2}\dot{u_3}\mW^{-1}_{u_3}S_2+\dot{y}_3\mW^{-1}_{y_3}T_1=z_4\mW_{x_4}\mW_{u_3}^{-1}-z_1\mW_{y_1}\mW_{y_3}^{-1}.
\end{equation}
Because $\mW_{y_4}=\mW_{y_2}$, and  $\mW_{y_1}=\mW_{y_3}$we can simplify \eqref{simp1} and \eqref{simp2}
\begin{align}
\label{z22}\dot{y}_4\mW^{-1}_{y_4}T_2&=z_4-z_2\\
\label{mult}\dot{u_3}\mW^{-1}_{u_3}S_2+\dot{y}_3\mW^{-1}_{y_3}T_1&=z_4\mW_{x_4}\mW_{u_3}^{-1}-z_1.
\end{align}
Then combining \eqref{z21} and \eqref{z22} eliminates $z_2$. 
\begin{equation}
\label{z42}\dot{x_2}\mW_{x_2}^{-1}S_1+\dot{y}_4\mW^{-1}_{y_4}T_2=z_4-z_1\mW_{u_1}\mW^{-1}_{x_2}.
\end{equation}
Multiplying \eqref{mult} by $(\mW_{x_4}\mW_{u_3}^{-1})^{-1}=\mW_{u_3}\mW_{x_4}^{-1}$ subtracting \eqref{z42} we have
\begin{equation}
\label{curvature}
\dot{z}(k^2)S_2-\dot{z}(k_1)S_1+\dot{z}(p^1)T_1-\dot{z}(p^2)T_2=z_1\left(\mW_{u_1}\mW^{-1}_{x_2}-\mW_{u_3}\mW_{x_4}^{-1}\right),
\end{equation}
where\footnote{Expressions are simplified using \eqref{idpath}}
\begin{align}
\dot{z}(k^1)&=\dot{x_2}\mW_{x_2}^{-1}=\dot{x_2}\left(V_0^{p^2}V_{p^2}^{k^1}\right)^{-1}=\dot{x}_2V^0_{k^2}\\
\dot{z}(k^2)&=\dot{u_3}\mW^{-1}_{x_4}=\dot{u_3}\left(U_0^{\om p^2}U_{\om p^2}^{k^2}\right)^{-1}=\dot{u}_3U^0_{k^2}\\
\dot{z}(p^1)&=\dot{y}_3\mW^{-1}_{y_3}\mW_{u_3}\mW_{x_4}^{-1}=\dot{y}_3\left(V_0^{p^1}\right)^{-1}\left(U_0^{\om p^1}V_{\om p^1}^{\om k^2}I^{k^2}\right)\left(U_0^{\om p^2}U_{\om p^2}^{k^2}\right)^{-1}=-\dot{y}_3V_{p^1}^0U^{\om p^1}_0V_{\om p^1}^0\label{zp1}\\
\dot{z}(p^2)&=\dot{y}_4\mW^{-1}_{y_4}=\dot{y}_4\left(V_0^{p^2}\right)^{-1}=\dot{y}_3V_{p^2}^0.
\end{align}
Now this is not quite the expression we initially guessed.   There is an extra term:\be \label{curious}z_1\left(\mW_{u_1}\mW^{-1}_{x_2}-\mW_{u_3}\mW_{x_4}^{-1}\right).\ee
We should also take note that the only $\dot{z}$ that is not related to a space-time coordinate by a simple parallel transport to the origin is the $\dot{z}$ which depends on $p^1$.  The  particular form appears again in the $W$ term.  To see this, we substitute in the values for the $W$'s as given in \cite{Gamma}.
\be
\left(\mW_{u_1}\mW^{-1}_{x_2}-\mW_{u_3}\mW_{x_4}^{-1}\right)=-V_0^{p^1}U_{p^1}^{\om k^1}I^{k^1}V^{p^2}_{k^1}V^0_{p^2}+U^{\om p^1}_0V_{\om p^1}^{\om k^2}I^{k^2}U_{k^2}^{\om p^2}U_{\om p^2}^0\label{u}
\ee
Using \eqref{gm1} and \eqref{idpath} we can rewrite this as 
\be
V_0^{p^1}U_{p^1}^{0}-U^{\om p^1}_0V_{\om p^1}^{0}\label{s}
\ee
Recall that 
\be
y_1=z_iW_{y_1}=z_iV_0^{p^1}
\ee
Then 
\be
z_1\left(\mW_{u_1}\mW^{-1}_{x_2}-\mW_{u_3}\mW_{x_4}^{-1}\right)=y_1U^0_{p_1}-y_1V_{p^1}^0U^{\om p^1}_0V_{\om p^1}^{0}
\ee
That is, the $W$ term is the difference between the parallel transport operator, transporting $y_1$ to the origin by the simple operator $U^0_{p^1}$ and a parallel transport identical to the one that brings $\dot{y}_1$ to the origin.  If we further manipulate this expression we find that 
\be
z_1\left(\mW_{u_1}\mW^{-1}_{x_2}-\mW_{u_3}\mW_{x_4}^{-1}\right)=y_1U^0_{p^1}(\delta-U_0^{p^1}V_{p^1}^0U^{\om p^1}_0V_{\om p^1}^{0})\label{cv}
\ee
The last term in \eqref{cv} now resembles the curvature terms discussed in section \ref{curved}.  To understand this,  we go back to the original picture.  Because the $\dot{z}$'s are elements of $T^*(0)$, the distances defined by $\dot{z}T_1$ etc. correspond to traveling along a path in the interaction space-time.  Thus \eqref{curvature}, implies that the right hand side correspond to a holonomy arising from tracing a closed curve from $z_1$ to $z_2$ to $z_4$ to $z_3$ and back to $z_1$.  In other words, there is some notion of curvature in the space-time at least in the interaction plane.  This curvature could explain the translation invariance imposed by the $z_1$ on the right hand side of \eqref{curvature}.  However, because we are working in a semiclassical limit with $G_N=0$, we do not believe that this curvature corresponds to the Einstein curvature tensor.  Moreover, it is likely that this "curvature" depends on the non-metricity, which is why it does not play a role in Einstein's general relativity.  

\subsection{A first order approximation}
To first order, however,
\begin{equation}
(\mW_{u_1}\mW^{-1}_{x_2})^a_b\approx(\mW_{u_3}\mW_{x_4}^{-1})^a_b\approx \delta^a_b+T^{ca}_bp^1.
\end{equation}
Thus the right hand side of the time delay equation vanishes and we are left with
\begin{equation}
\dot{z}(k^2)S_2-\dot{z}(k_1)S_1=\dot{z}(p^2)T_2-\dot{z}(p^1)T_1.
\end{equation}
From this we can determine an expression for the time delay to first order.  Expanding the parallel transport operators to first order and substituting in the velocity relationship given by the equations of motion:
\begin{align}
\dot{z}(k^1)&=\frac{1}{2m_{k^1}}\frac{\partial D^2(k^1)}{\partial k^1}V_{k^1}^0\qquad
\dot{z}(k^2)=\frac{1}{2m_{k^2}}\frac{\partial D^2(k^2)}{\partial k^2}U_{k^2}^0\\
\dot{z}(p^1)&=\frac{1}{2m_{p^1}}\frac{\partial D^2(p^1)}{\partial p^1}U_{p^1}^0\qquad
\dot{z}(p^2)=\frac{1}{2m_{p^2}}\frac{\partial D^2(p^2)}{\partial p^2}V_{p^2}^0
\end{align}
Notice that each of the velocities depend on only one momentum.  We should also point out that, while the velocities are functions of the momenta, they still live in $T^*(0)$; they are fundamentally a space-time object.  Now if we assume that the emitter and detector are at rest with respect to each other as seen by the interaction plane $\left(\dot{z}(k^1)=\dot{z}(k^2)\right)$, then the expression above is reduced to
\begin{equation}
\label{want}(S_2-S_1)\dot{z}(k^2)=\dot{z}(p^2)T_2-\dot{z}(p^1)T_1.
\end{equation}
However, if $\dot{z}(k^1)=\dot{z}(k^2)$ then we must have that 
\begin{equation}
\label{same}\frac{1}{m_{k^1}}\frac{\partial D^2(k^1)}{\partial k^1}V_{k^1}^0=\frac{1}{m_{k^2}}\frac{\partial D^2(k^2)}{\partial k^2}U_{k^2}^0.
\end{equation}
 We can expand the parallel transport operators to first order as 
\be
\label{transport}(U^k_r)_a^b=\delta^b_a-\Gamma^{cb}_a(r-k)_c\qquad(V^k_r)_a^b=\delta^b_a-\Gamma^{bc}_a(r-k)_c
\ee
Thus to first order,  $U^k_r$ and $V^k_r$ differ only when torsion is present.  If the torsion is zero, then the connection coefficients are symmetric, which means $U^k_r=V^k_r$.  Then simply taking $k^1=k^2$ satisfies the relationship \eqref{same}.  Otherwise, for the two detectors to appear at rest with respect to each other in the interaction plane, they cannot have the same momentum.  This is due to the phenomenon called dual gravitational lensing, which will be discussed in Section \ref{dual}.  \\\\
Now we can proceed with the calculation of the time delay.  If we normalize $\dot{z}(k^i)$ so that 
\be
|\dot{z}(k^i)|\equiv \sqrt{(\dot{z}(k^1))^2}=1,
\ee 
then  \eqref{want} becomes
\be
\label{want2}(S_2-S_1)\hat{K}=\dot{z}(p^2)T_2-\dot{z}(p^1)T_1,
\ee
 where
\begin{equation}
\hat{K}^c\equiv\frac{\dot{z}(k^1)^c}{|\dot{z}(k^1)|}=\frac{\dot{z}(k^2)^c}{|\dot{z}(k^2)|} \footnote{recall that we have assumed that $\dot{z}(k^1)=\dot{z}(k^2)$}.\end{equation} 
Now we want decompose the right hand side of \eqref{want2} into its component in the $\hat{K}$ direction and its component perpendicular to $\hat{k}$, which we will call $\hat{R}$.  
\begin{equation}
\label{prop}
\dot{z}^a(p^i)=(\hat{K}\cdot \dot{z}(p^i))\hat{K}^a+\sqrt{(\hat{K}\cdot \dot{z}(p^i))^2-(\dot{z}(p^i))^2}\hat{R}^a,
\end{equation}
From \eqref{want2} we know that 
\begin{align}
\label{1}
\Delta S=(\hat{K}\cdot \dot{z}(p^2))T_2-(\hat{K}\cdot \dot{z}(p^1))T_1,\end{align}
and
\begin{align}
\label{2}
T_2\sqrt{(\hat{K}\cdot \dot{z}(p^i))^2-(\dot{z}(p^2))^2}-T_1\sqrt{(\hat{K}\cdot \dot{z}(p^i))^2-(\dot{z}(p^1))^2}=0.
\end{align}
Combining \eqref{1} and \eqref{2} we have that 
\begin{equation}
\Delta S=\left[(\hat{K}\cdot \dot{z}(p^2))-\sqrt{(\hat{K}\cdot \dot{z}(p^i))^2-(\dot{z}(p^2))^2}\right]T_2-\left[(\hat{K}\cdot \dot{z}(p^1))-\sqrt{(\hat{K}\cdot \dot{z}(p^i))^2-(\dot{z}(p^1))^2}\right]T_1
\end{equation}
To first order, this is approximately
\be
\label{time}
\Delta S\approx \frac{(\dot{z}(p^2))^2}{2\hat{K}\cdot \dot{z}(p^2)}T_2-\frac{(\dot{z}(p^1))^2}{2\hat{K}\cdot \dot{z}(p^1)}T_1
\ee
We know have a coordinate independent expression for the time delay calculation to first order.  However, in order to actually compute the time delay we would need to choose a coordinate system.  

\section{ \label{MM}Time Delay in Connection Normal Coordinates}
\subsection{Defining the connection normal coordinates}
To proceed, we need to define a coordinate system.  For all coordinate systems, we will take the convention that $\eta=(+---)$.  A common and often very convenient choice of coordinates are the Riemann normal coordinates, which are defined by insisting that at some point $k$ (usually the origin)
\begin{equation}
g^{\mn}(k)=\eta^{\mn}
\qquad\mathrm{and}\qquad g^{\mn,a}(k)=0
\end{equation}
and thus the connection is completely defined by the torsion and non-metricity tensors.  This is the coordinate system used in \cite{Gamma}.  However, there is another important normal coordinate system, which we refer to as connection normal coordinates.  We will calculate the time delay in connection normal coordinates to provide a check to the claim that the calculation is coordinate independent and to see if there is any underlying physics that is not readily apparent in the Riemann normal coordinates.  As with the Riemann normal coordinates, we will require that $g^{\mn}(k)=\eta^{\mn},$ but instead of setting the derivatives of the metric tensor to 0 at $k$, we will require that the symmetric part of the connection be zero.  That is 
\be\Gamma^{(ab)}_c(k)=0,\ee which means \be\Gamma^{ab}_c=\Gamma^{[ab]}_c=\frac{1}{2}T^{ab}_c.\ee  Thus we have that the connection is equal to the torsion.  This choice then fixes the first derivative of the metric tensor.  To see this 
\begin{align}
\Gamma^{(ab)i}+\Gamma^{(ai)b}=g^{bi,a}+\frac{1}{2}\left(T^{iab}+T^{bai}-2N^{abi}\right)
\end{align}
Setting the symmetric part of the connection coefficient to zero we have that 
\begin{equation}
g^{bi,a}=-\frac{1}{2}\left(T^{iab}+T^{bai}-2N^{abi}\right)
\end{equation}
Let us define 
\be\label{N}\mathcal{N}^{abi}\equiv N^{abi}-\frac{1}{2}\left(T^{iab}+T^{bai}\right).\ee  
Then 
\begin{equation}
g^{bi,a}=\mathcal{N}^{abi}.
\end{equation}
In the case where the metric is compatible, the Riemann normal coordinates and the connection normal coordinates are the same.  It is only when we introduce non-metricity that there is a difference. \\\\
Since we are only considering momenta close to the origin, we can Taylor expand the metric tensor about the origin.  To first order, 
\begin{equation}
g^{ab}(p)=\eta^{ab}+g^{ab,c}p_c+O(p^2)=\eta^{ab}+\mathcal{N}^{cab} p_c+O(p^2).  
\end{equation}
 From this we can construct  the distance function from the origin to a point in $\mathcal{P}$.  In the Riemann normal coordinates,
\begin{equation}
D^2(p,0)=\eta^{ab}p_ap_b=E^2-\vec{p}^2=m^2,
\end{equation}
which is the standard relationship in relativity between the total energy, the kinetic energy and the rest mass.  However, in connection normal coordinates, 
\begin{equation}
m^2=D^2(p,0)=\eta^{ab}p_ap_b +\mathcal{N}^{cab}(0) p_c p_a p_b=E^2-\vec{p}^2+\mathcal{N}^{cab}(0) p_c p_a p_b.  
\end{equation}
Thus the mass is dependent on energy and the non-metricity of momentum space.  Moreover, the "measured mass" is not the same as the mass one would obtain by taking $p_0^2-p^ap_a=m^2$.  This is the expected dispersion relationship that we expect to see.  Notice that if the units are consistent, then $\mathcal{N}$ must have the units of $\frac{1}{[p]}=\frac{1}{m}$.  
We will now compute the time delay in connection normal coordinates.  First we need to calculate $\dot{x}$.  Recall that in connection normal coordinates 
\begin{equation}
D^2(k)=\eta^{ab}k_ak_b+\mN^{cab}(0)k_ck_ak_b.
\end{equation}
Plugging this into the equation for the velocity, 
\begin{equation}
\label{x}\dot{x}^c=\frac{1}{2m}\{2k^c+\left(\mN^{c ab}(0)+\mN^{ac b}(0)+\mN^{ abc}(0)\right)k_ak_b\}
\end{equation}
For convenience, we define
 \be\label{M}\mathcal{M}^{abc}\equiv \frac{1}{2}\left(\mN^{abc}+\mN^{bac}+\mN^{bca}\right).  
 \ee
 Then \eqref{x} can be written as
\begin{equation}
\dot{x}^c=\hat{k}^c+m\mathcal{M}^{cab}\hat{k}_a\hat{k}_b.
\end{equation}
where we have defined $\hat{k}=\frac{k}{m}$.  Substituting the definition of $\mathcal{N}$ given by \eqref{N} into \eqref{M}
\be
\mathcal{M}^{abc}=\frac{1}{2}\left(N^{abc}+N^{bac}+N^{bca}\right)-\frac{1}{4}\left(T^{cab}+T^{abc}+2T^{cba}\right).
\ee
Thus 
\be
\label{M1}\mathcal{M}^{abc}\hat{k}_b\hat{k}_c=\frac{1}{2}\left(N^{abc}+N^{bac}+N^{bca}\right)\hat{k}_b\hat{k}_c-\frac{1}{4}\left(T^{cab}+T^{abc}+2T^{cba}\right)\hat{k}_b\hat{k}_c
\ee
We can rewrite 
\be
\label{T}T^{dab}\hat{k}_b\hat{k}_d=T^{bad}\hat{k}_b\hat{k}_d.\ee
Then, using \eqref{T} and the fact that $T^{bda}$ is antisymmetric in $b$ and $d$, \eqref{M1} becomes
\be
\mathcal{M}^{abc}\hat{k}_b\hat{k}_c=\frac{1}{2}\left(N^{abc}+N^{bac}+N^{bca}\right)\hat{k}_b\hat{k}_c.  
\ee
This show that the definition of $\dot{x}$ does not have any dependence on torsion.  \\\\
Expanding the parallel transport operators using \eqref{transport} and calculating the velocities using the procedure outlined above, we have that

\begin{align}
\dot{z}(k^1)^c&\approx\left( \hat{k}_1^a+m_{k^1}\mathcal{M}^{abd}\hat{k}^1_b\hat{k}^1_d\right)\left(\delta^c_a-\frac{1}{2}T^{cb}_ak^1_b\right)\approx\hat{k}_1^a+m_{k^1}\mathcal{M}^{abd}\hat{k}^1_b\hat{k}^1_d-\frac{1}{2}T^{cba}k^1_bk^1_a
\\
\dot{z}(k^2)^c&\approx\left( \hat{k}_2^a+m_{k^2}\mathcal{M}^{abd}\hat{k}^2_b\hat{k}^2_d\right)\left(\delta^c_a-\frac{1}{2}T^{d c}_ak^2_d\right)\approx\hat{k}_2^a+m_{k^2}\mathcal{M}^{abd}\hat{k}^2_b\hat{k}^2_d-\frac{1}{2}T^{bca}k^2_bk^a_a
\\\label{p2}
\dot{z}(p^1)^c&\approx\left(\hat{p}_2^a+E_{p^2}\mathcal{M}^{abd}\hat{p}^2_b\hat{p}^2_d\right)\left(\delta^c_a-\frac{1}{2}T^{bc}_ap^2_b\right)\approx\hat{p}_2^a+E_{p^2}\mathcal{M}^{abd}\hat{p}^2_b\hat{p}^2_d-\frac{1}{2}T^{bca}p^2_bp^2_a
\\
\label{p1}\dot{z}(p^2)^c&\approx\left(\hat{p}_1^a+E_{p^1}\mathcal{M}^{abd}\hat{p}^1_b\hat{p}^1_d\right)\left(\delta^c_a-\frac{1}{2}T^{cb}_ap^1_b\right)\approx\hat{p}_1^a+E_{p^1}\mathcal{M}^{abd}\hat{p}^1_b\hat{p}^1_d-\frac{1}{2}T^{cba}p^1_bp^1_a
\end{align}
We now need to calculate the projection of $\dot{z}(p^i)$ onto $\hat{K}$. To first order 
\begin{align}
\label{k1}\hat{K}\cdot \dot{z}(p^1)&=1+m_{k^1}\mathcal{M}^{cbd}\left(\hat{k}^1_c\hat{p}^1_b\hat{p}^1_d+\hat{k}^1_b\hat{k}^1_d\hat{p}^1_c-\hat{k}^1_b\hat{k}^1_d\hat{k}^1_c\right)-\frac{1}{2}T^{cba}\hat{k}^1_b\hat{k}^1_a\hat{p}^1_c-\frac{1}{2}T^{cba}\hat{p}^1_b\hat{p}^1_a\hat{k}^1_c\\
\label{k2}\hat{K}\cdot \dot{z}(p^2)&=1+m_{k^1}\mathcal{M}^{cbd}\left(\hat{k}^1_c\hat{p}^2_b\hat{p}^2_d+\hat{k}^1_b\hat{k}^1_d\hat{p}^2_c-\hat{k}^1_b\hat{k}^1_d\hat{k}^1_c\right)-\frac{1}{2}T^{cba}\hat{k}^1_b\hat{k}^1_a\hat{p}^2_c-\frac{1}{2}T^{bca}\hat{p}^2_b\hat{p}^2_a\hat{k}^1_c.
\end{align}
 Finally, we need to calculate
\be
(\dot{z}(p^i))^2=2E_{p_i}\mathcal{M}^{cbd}\hat{p}^i_b\hat{p}^i_d\hat{p}^i_c.
\ee
Plugging this into \eqref{time}, we have that to first order
\begin{equation}
\label{deltas}
\Delta S\approx \frac{(\dot{z}(p^2))^2}{2\hat{K}\cdot \dot{z}(p^2)}T_2-\frac{(\dot{z}(p^1))^2}{2\hat{K}\cdot \dot{z}(p^1)}T_1\approx E_{p_2}\mathcal{M}^{cbd}\hat{p}^2_b\hat{p}^2_d\hat{p}^2_cT_2-E_{p_1}\mathcal{M}^{cbd}\hat{p}^1_b\hat{p}^1_d\hat{p}^1_cT_1.
\end{equation}
Note that the torsion term from \eqref{k1} and \eqref{k2} completely drops out of the time delay calculation to first order and we are only left with terms involving $\mathcal{M}^{cbd}\hat{p}^i_c\hat{p}^i_b\hat{p}^i_d$, which we know from Section \ref{MM},  only depends on non-metricity.  \\\\
Now to recover the results from \cite{Gamma}.  Suppose the energy of the first photon is much less than the energy of the second one.  For simplicity we will  take the torsion to be zero, as they do in \cite{Gamma}, so that we can assume that $\hat{p}^1\approx\hat{p}^2$.  However, in the next section we will show why this assumption is unnecessary. Then 
\begin{equation}
\label{approxs}
\Delta S=T_2E_2\mathcal{M}^{cbd}\hat{p}_b\hat{p}_d\hat{p}_c.
\end{equation}
 Now comparing this to the results derived in \cite{Gamma}, we find that 
 \begin{equation}
 \frac{1}{2}\tilde{N}^{cbd}=\mathcal{M}^{cbd}=\frac{1}{2}\left(N^{cbd}+N^{bcd}+N^{bdc}\right),
 \end{equation}
where $\tilde{N}^{cbd}$ is the non-metricity tensor in Riemann normal coordinates. Thus we find that the time delay calculated here is consistent with the one found in \cite{Gamma}.  \\\\ 

\subsection{Dual Gravitational Lensing}\label{dual}
As we mentioned in Section 5, if torsion is present then the direction of the motion of the particle emitted in parallel will not necessarily be parallel when they reach the detector.  However, from \eqref{prop} we know that 
$\dot{z}(p^1)$ and $\dot{z}(p^1)$ propagate in the same plane spanned by $\hat{K}$ and $\hat{R}$, so  
\begin{align}
\dot{z}^a(p^i)&=\hat{K}\cdot \dot{z}(p^i)(\hat{K}+\hat{R})^a-\frac{(\dot{z}(p^i))^2}{2\hat{K}\cdot \dot{z}(p^i)}\hat{R}^a\\
\label{prop2}&\approx \hat{K}\cdot \dot{z}(p^i)(\hat{K}+\hat{R})^a-E_{p_i}\mathcal{M}^{cbd}\hat{p}^i_b\hat{p}^i_d\hat{p}^i_c\hat{R}^a.
\end{align}
If we take the non-metricity to be zero then \eqref{prop2} reduces to 
\be
\dot{z}(p^i)=\hat{K}\cdot \dot{z}(p^i)(\hat{K}+\hat{R}).
\ee
We will define the direction of propagation in the interaction place as $e^+$.  If we normalize $\dot{z}(p^i)$ with respect to $\hat{K}$, then we can simply write $\dot{z}(p^i)^a=e^+$.  We will denote the covector by $e_-^b=e^+_a\eta^{ab}$.
But much like with the detectors, if the photons appear to be traveling parallel in the interaction plane, they are not emitted parallel.  We now want to calculate to what extent they are different.  To see this we invert the relationships given by \eqref{p1} and \eqref{p2} with the non-metricity taken to be zero.
\begin{equation}
\hat{p}^a_1\approx e^a_--\frac{E_1}{2}T^{a+}_-
\end{equation}
\begin{equation}
\hat{p}^a_2\approx e^a_-+\frac{E_2}{2}T^{a+}_-
\end{equation}
where we have denoted $T^{ab}_ce^c_-e^+_b=T^{a+}_-$. Thus 
\begin{equation}
\label{diff}\hat{p}^a_2-\hat{p}^a_1=\frac{E_1+E_2}{2}|T^+_-|\hat{T}
\end{equation}
Where $\hat{T}$ is perpendicular to the  propagation vector $e^+$ and thus we can think of the difference between $\hat{p}_2$ and $\hat{p}_1$ as being a rotation in the plane perpendicular to the propagation by and angle $\theta=\frac{E_1+E_2}{2}|T^+_-|$.   We we have defined
 \be
 |T^+_-|\equiv\sqrt{T^{c+}_-\eta_{cd}T^{d+}_-}\quad \mathrm{and} \quad\hat{T}=\frac{T^{a+}_-}{|T^+_-|}.\ee 
 Using equation \eqref{diff} we can write
 \be
 \hat{p}^a_2=\hat{p}^a_1+\frac{E_1+E_2}{2}|T^+_-|\hat{T}
 \ee
 Plugging this into \eqref{deltas} we find that 
 \be
  E_{p_2}\mathcal{M}^{cbd}\hat{p}^2_b\hat{p}^2_d\hat{p}^2_cT_2-E_{p_1}\mathcal{M}^{cbd}\hat{p}^1_b\hat{p}^1_d\hat{p}^1_cT_1\approx\left(E_{p_2}T_2-E_{p_1}T_1\right)\mathcal{M}^{cbd}\hat{p}^1_b\hat{p}^1_d\hat{p}^1_c +\mathcal{O}(\Gamma,T)
\ee
Applying the assumption that $E_2>>E_1$ we recover the time delay given by \eqref{approxs}\\\\
 We would expect to find a similar angle of rotation for $\hat{k}^1$ and $\hat{k}^2$ given by 
 \begin{equation}
 \label{rotate}\theta=\frac{m_1+m_2}{2}|T^+_-|\hat{T}.
 \end{equation} 
 In Section 5, we noted that in the presence of torsion the assumption that $\dot{z}(k^1)$ was parallel to $\dot{z}(k^2)$ meant that $k^1$ could not be parallel to $k^2$.  We now see from \eqref{rotate}that the reason for this is that the torsion generates a rotation as the velocities are parallel transported to the interaction plane that depends on the masses of the particles.

 \section{Discussion}
 Although we recovered the same results as \cite{Gamma}, the approach taken here differs significantly in a couple of ways.  The same general approach was taken to find the time delay.  However, we chose to determine the time delay using the velocity, defined by the proper time derivative of the space coordinates rather than using the fact that in Riemann normal coordinates the velocity is given by the momentum divided by the mass.  By doing so we are able to derive a coordinate independent expression for the time delay.  In addition, the use of the velocities, the $\dot{z}$'s, removes some of the ambiguity of the upper case momenta ($K^i$ and $P^i$) used in \cite{Gamma}, which are interpreted as the physical momenta ($k^1$ for example) parallel transported to the origin.  However, the meaning of parallel transporting an element of the manifold to the origin is unclear.
\\
Another reason for this approach is that it makes the impact of the geometry of momentum space on the emergence of space-time clearer.   In Section 2 we argue that the energy plays an important role in the construction of quantum geometry via the energy dependence of the light signal traveling between clocks.  In order to see how the energy dependence manifests in the emergence of space-time, we needed to have some way of expressing distance as the integral of velocity over time.  In this case we assumed the velocity was constant.  In the approach taken in \cite{Gamma}, the concept of distance in space-time is blurred by the use of the momentum in place of velocity.  By working with velocities which are strictly elements of the cotangent planes, we see that the energy dependence does arise from the energy dependence of the velocity associated with the particle's momentum (the $\dot{x}$'s). \\\\

\section{Conclusion} 
In this paper, we discussed the appearance of curvature like terms due to the difference in left and right parallel transport operators and their relation to the unexpected term that arose in the time delay calculation \eqref{curious}. We also derived a general expression for the time delay to first order, which is given by
 \begin{equation*}
 \Delta S\approx \frac{(\dot{z}(p^2))^2}{2\hat{K}\cdot \dot{z}(p^2)}T_2-\frac{(\dot{z}(p^1))^2}{2\hat{K}\cdot \dot{z}(p^1)}T_1
 \end{equation*}
 and the specialized to connection normal coordinates.  In this coordiante system we found the expected 
 dispersion relationship
\begin{equation*}
m^2=E^2-\vec{p}^2+\mathcal{N}^{cab}(0) p_c p_a p_b.  
\end{equation*}
   The time delay in connection normal coordinates is given by
 \begin{equation*}
 \Delta S\approx E_{p_2}\mathcal{M}^{cbd}\hat{p}^2_b\hat{p}^2_d\hat{p}^2_cT_2-E_{p_1}\mathcal{M}^{cbd}\hat{p}^2_b\hat{p}^2_d\hat{p}^2_cT_1.
 \end{equation*}
 After making the assumption that $E_2>>E_1$,  we find that this reduces to 
 \begin{equation*}
 \Delta S=T_2E_2\mathcal{M}^{cbd}\hat{p}_b\hat{p}_d\hat{p}_c,
 \end{equation*}
 which is the same result as in \cite{Gamma}.  Thus we have that the time delay is not effected by the choice of coordinate system.  We have also shown that the time delay is not impacted by by the presence of torsion.  \\\\
 It should be noted that this is true only for the particular choice of energy-momentum conservation law.  In recent work by Oliveira \cite{Ricky}, it is shown that to first order, the choice of conservation law for the interactions does not effect the results if there is no torsion.  However, when torsion is present, for certain choices, the right hand side of \eqref{curvature} does not vanish and the translational invariance is broken.  As mentioned in Section 7,  a similar problem is expected to arise when the curvature is taken to be non-zero.  We also expect that there will be a number of other interesting phenomena occurring when the curvature of momentum space is non-zero which deserve further study.
\section{Acknowledgements}
First and foremost, I would like to sincerely thank  my advisor, Lee Smolin,  for all of his guidance, encouragement and support during this project.   I would also like to thank Laurent Freidel and Jos\'e Ricardo Oliveira for all of their comments and suggestions. The funding for this project was provided by the Perimeter Institute of Theoretical Physics and the University of Waterloo.

\end{document}